\documentclass{mn2e}

\usepackage{epsfig}

\usepackage{mathptm}

\title[9C continued: a radio-source survey at 15 GHz]
  {9C continued: results from a deeper radio-source survey at 15~GHz}
\author[Waldram et al.]
{E.~M.~Waldram,$^1$\thanks{Email: e.m.waldram@mrao.cam.ac.uk}
 G.~G.~Pooley,$^1$
 M.~L.~Davies,$^1$
 \newauthor
 K.~J.~B.~Grainge$^{1,2}$
 and P.~F.~Scott$^1$\\
 $^1$Astrophysics Group, Cavendish Laboratory, J J Thomson Avenue, Cambridge, CB3 0HE\\
 $^2$Kavli Institute for Cosmology Cambridge, Madingley Road, Cambridge, CB3 0HA}
\date{Accepted ????.  Received ????.}
\pagerange{\pageref{firstpage}--\pageref{lastpage}}
\pubyear{2009}

\begin{document}

\maketitle

\label{firstpage}

\begin{abstract}
The 9C survey of radio sources with the Ryle telescope at 15.2~GHz was set up to survey the fields of the cosmic microwave background telescope, the Very Small Array. In our first paper we described three regions of the survey, constituting a total area of 520~deg$^{2}$ to a completeness limit of $\approx25$~mJy. Here we report on a series of deeper regions, amounting to an area of 115~deg$^{2}$ complete to $\approx10$~mJy and of 29~deg$^{2}$ complete to $\approx5.5$~mJy. We have investigated the source counts and the distributions of the 1.4~to~15.2~GHz spectral index ($\alpha_{1.4}^{15.2}$) for these deeper samples. The whole catalogue of 643~sources is available online.

Down to our lower limit of 5.5~mJy we detect no evidence for any change in the differential source count from the earlier fitted count above 25~mJy, $n(S) = 51(S/{\rm Jy})^{-2.15}\, {\rm Jy}^{-1}{\rm sr}^{-1}$.

We have matched both our new and earlier catalogues with the NRAO VLA Sky Survey (NVSS) catalogue at 1.4~GHz. For samples of sources selected at 15.2~GHz, in three flux density ranges, we detect a significant shift in the median value of $\alpha_{1.4}^{15.2}$; samples with higher flux densities have higher proportions of sources with flat and rising spectra. We suggest that this observed shift is consistent with a model containing two distinct source populations having differently sloped source counts.  Samples selected at 1.4~GHz contain significantly smaller proportions of sources with flat and rising spectra. Also, in our area complete to $\approx10$~mJy, we find 5 sources between 10~to~15 mJy, amounting to 4.3~per~cent of sources in this range, with no counterpart in the NVSS catalogue. These results illustrate the problems inherent in using a low frequency catalogue to characterise the source population at a much higher frequency and emphasise the value of our blind 15.2-GHz survey.
\end{abstract}

\begin{keywords}
surveys -- cosmic microwave background -- radio continuum: general -- galaxies:evolution
\end{keywords}

\section{Introduction}
As explained in our first paper (Waldram et al. 2003), hereafter Paper I, the 9C survey with the Ryle Telescope (RT) at 15.2~GHz was designed specifically as part of the observing strategy of the Very Small Array (VSA) (Watson et al. 2003). Foreground radio sources are a major contaminant in the cosmic microwave background (CMB) observations of the VSA at 34~GHz and the prime purpose of the 9C survey has been to provide a catalogue of radio sources in each of the VSA fields (Taylor et al. 2003; Cleary et al. 2005, 2008). This has meant that the positions of the 9C fields and their depth in different areas have been determined by the VSA observing programme. For the VSA compact array ($\ell = $~150~to~800) we covered a total area of 520~deg$^{2}$ to a completeness limit of 25~mJy but for the extended array ($\ell = $~300~to~1500) and now the super-extended array ($\ell = $~600~to~2500) deeper surveying over some smaller regions has been required. It is these regions which are described here.

In addition to its specific application to the VSA observations, the 9C survey has proved to be of much wider interest. There have been other blind surveys at high radio frequencies, notably with the Australia Telescope Compact Array (ATCA) at 18 and 20~GHz (Ricci et al. 2004; Sadler et al 2006; Massardi et al. 2008), but 9C was the first such survey of any extent and is the only one to reach such low flux densities. It provides an important opportunity for exploring the radio-source population at high frequencies. Samples of 9C sources have been followed up with simultaneous multifrequency observations at 1.4, 4.8, 15.2, 22 and 43~GHz (Bolton et al. 2004) and there have been two further papers based on these data: Bolton et al. 2006a, which describes results from 5-GHz MERLIN and VLBA observations, and Bolton et al. 2006b, which examines variability at 15.2~GHz.

As a blind survey at 15.2~GHz, 9C is particularly valuable in the study of source counts at high radio frequencies. Point sources are a significant foreground in the CMB observations of many instruments (e.g. the \textit{Planck Surveyor}) and, even if the brighter sources can be identified and subtracted from the data, there remains the problem of residual confusion from the fainter, unsubtracted source-population. This effect can be evaluated only from a knowledge of the source counts below the subtraction limit.  Blind surveys at frequencies of 30~GHz and above are extremely problematical because of the small field of view of the available telescopes, so one approach has been to identify sources from a lower frequency survey, with a measured source count, and follow them up at the higher frequency. 9C, at 15.2~GHz, is comparatively close to the target frequencies and, therefore, has a considerable advantage over lower-frequency surveys.

We have used this method to make empirical estimates of the source counts at higher frequencies (including those in the lower \textit{Planck} bands) by taking the spectral index distributions from the Bolton data together with our known 15.2-GHz source count (Waldram et al. 2007). Comparison with the theoretical model of de Zotti et al. (2005) shows good agreement below 43~GHz but at higher frequencies there is some divergence towards higher flux densities, indicating that our predictions imply fewer flat spectrum sources. However, the results of Sadler et al. (2008) who have made simultaneous observations with ATCA of a sample of sources at 20 and 95~GHz, do appear to confirm our own estimates of the count at 90~GHz in the common flux-density range.

More recently, Mason et al. (2009) have made a 31-GHz survey of the fields of the Cosmic Background Imager (CBI, Mason et al. 2003) by following up known extragalactic sources from the NVSS survey at 1.4~GHz (Condon et al. 1998). They have estimated the spectral index distribution from 1.4 to 31~GHz and hence the 31-GHz source count in the range 0.5 to 10~mJy, which has enabled them to make an assessment of the contribution of unresolved point sources to the CBI power spectrum.  Our results here provide a useful comparison with their work.

Also, a subset of the sources in our present catalogue has now been followed up at 30~GHz with observations made using the 32-m, Toru\'{n} telescope (Gawro\'{n}ski et al., in preparation).

In this paper, we first explain the choice of fields and our methods of observation and data analysis, together with estimates of completeness (Sections 2, 3 and 4). In Section 5 we present the extended source counts and in Section 6 investigate the correlation of our source lists with the NVSS survey. Section 7 describes the source catalogue. Section 8 discusses our conclusions and Section 9 our plans for future work. 

\section{The survey areas}

\begin{figure}
        \centerline{\epsfig{file=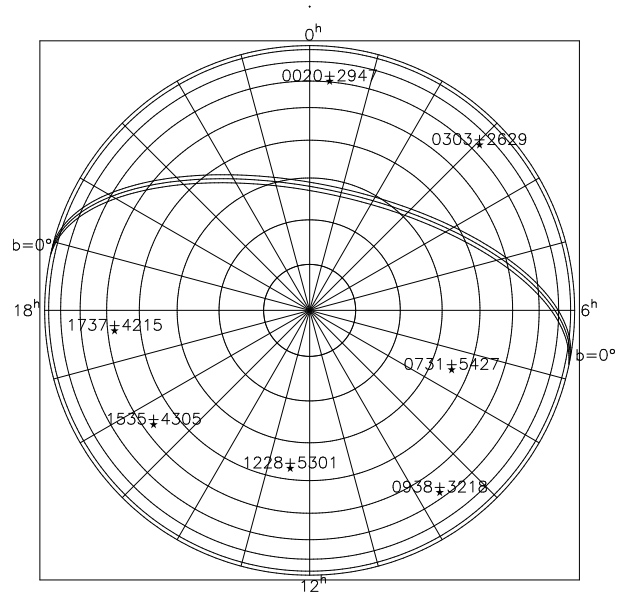,
        angle=270,
        width=8.0cm,clip=}}
        \caption{The deeper 9C survey areas: an equatorial plane projection with the N. pole at the centre. The Declination circles are at intervals of $10^\circ$ and the Galactic plane is shown.}
\label{fig:fields eqproj}
\end{figure} 

\begin{table}
\caption{The areas complete to $\approx10$ mJy}
\begin{tabular}{ccccccc}
\hline
 Field && RA ($^{\rm h}\ ^{\rm m}\ ^{\rm s}$) to RA ($^{\rm h}\ ^{\rm m}\ ^{\rm s}$) && Dec ($^\circ\ ^\prime\ ^{\prime\prime}$) to Dec ($^\circ\ ^\prime\ ^{\prime\prime}$)  \\
\hline
 0020+2947  &&  00 06 48.2 \space\space 00 33 04.6  &&  +26 37 55 \space\space+32 55 20  \\
 0303+2629  &&  02 49 36.5 \space\space 03 16 03.3  &&  +24 26 19 \space\space+28 32 04  \\
 0731+5427  &&  07 22 05.0 \space\space 07 40 01.2  &&  +53 09 29 \space\space+55 44 20  \\
 0938+3218  &&  09 29 38.0 \space\space 09 46 08.0  &&  +30 43 45 \space\space+33 51 55  \\
 1228+5301  &&  12 17 08.5 \space\space 12 38 43.9  &&  +51 27 16 \space\space+54 33 45  \\
 1535+4305  &&  15 22 16.7 \space\space 15 47 10.1  &&  +40 43 16 \space\space+45 25 52  \\
 1735+4215  &&  17 29 40.3 \space\space 17 43 57.2  &&  +41 13 24 \space\space+43 17 30  \\
\hline
\end{tabular}
\label{table:areastotal}
\end{table}

\begin{table}
\caption{The areas complete to $\approx5.5$ mJy}
\begin{tabular}{ccccccc}
\hline
 Field && RA ($^{\rm h}\ ^{\rm m}\ ^{\rm s}$) to RA ($^{\rm h}\ ^{\rm m}\ ^{\rm s}$) && Dec ($^\circ\ ^\prime\ ^{\prime\prime}$) to Dec ($^\circ\ ^\prime\ ^{\prime\prime}$)  \\
\hline
0020+2947 && 00 11 55.7\space\space 00 20 08.0 && +27 09 39.0\space\space +29 16 04 \\
          && 00 20 08.0\space\space 00 22 45.5 && +28 43 43.0\space\space +29 16 04 \\
          && 00 25 00.8\space\space 00 30 54.6 && +31 53 24.0\space\space +32 24 07 \\
\hline
0303+2629 && 02 55 27.2\space\space 02 58 26.5 && +24 56 03.0\space\space +25 27 56 \\
          && 02 58 26.5\space\space 03 01 23.7 && +25 27 56.0\space\space +25 58 27 \\
          && 02 58 21.7\space\space 03 04 22.8 && +26 29 55.0\space\space +26 59 31 \\
\hline
0731+5427 && 07 22 05.0\space\space 07 31 04.6 && +53 09 29.0\space\space +53 40 27 \\
          && 07 26 33.2\space\space 07 40 01.2 && +53 40 27.0\space\space +55 13 28 \\
          && 07 22 05.0\space\space 07 31 17.7 && +55 13 28.0\space\space +55 44 20 \\
\hline
0938+3218 && 09 29 38.0\space\space 09 40 46.9 && +30 43 45.0\space\space +32 49 53 \\
          && 09 43 01.8\space\space 09 46 08.0 && +31 14 44.0\space\space +31 46 05 \\
          && 09 40 46.9\space\space 09 43 35.0 && +32 17 15.0\space\space +32 49 55 \\
          && 09 32 08.6\space\space 09 35 23.3 && +32 49 53.0\space\space +33 21 31 \\
          && 09 37 38.7\space\space 09 40 46.9 && +32 49 53.0\space\space +33 21 16 \\
\hline
1228+5301 && 12 21 19.2\space\space 12 25 44.9 && +51 58 14.0\space\space +52 30 36 \\
          && 12 25 44.9\space\space 12 30 01.0 && +51 58 14.0\space\space +54 02 56 \\
          && 12 34 22.2\space\space 12 38 48.0 && +51 58 21.0\space\space +52 30 43 \\
          && 12 20 54.5\space\space 12 25 44.9 && +53 31 51.0\space\space +54 02 56 \\
\hline
1535+4305 && 15 34 28.4\space\space 15 44 07.5 && +40 43 16.0\space\space +42 49 11 \\
          && 15 44 07.5\space\space 15 47 10.1 && +42 18 13.0\space\space +42 49 11 \\
          && 15 22 16.7\space\space 15 28 44.7 && +42 48 44.0\space\space +43 19 48 \\
          && 15 22 16.7\space\space 15 25 32.3 && +43 19 48.0\space\space +43 51 07 \\
          && 15 40 59.3\space\space 15 44 46.9 && +44 23 59.0\space\space +44 55 40 \\
          && 15 24 45.0\space\space 15 31 50.2 && +44 54 51.0\space\space +45 25 52 \\
\hline
1735+4215 && 17 29 40.3\space\space 17 43 57.2 && +41 13 24.0\space\space +42 15 51 \\
          && 17 29 40.3\space\space 17 33 16.0 && +42 15 51.0\space\space +42 46 40 \\
          && 17 36 50.1\space\space 17 43 57.2 && +42 15 51.0\space\space +42 46 40 \\
\hline
\end{tabular}
\label{table:areasdeep}
\end{table}

\begin{figure*}
        {\epsfig{file=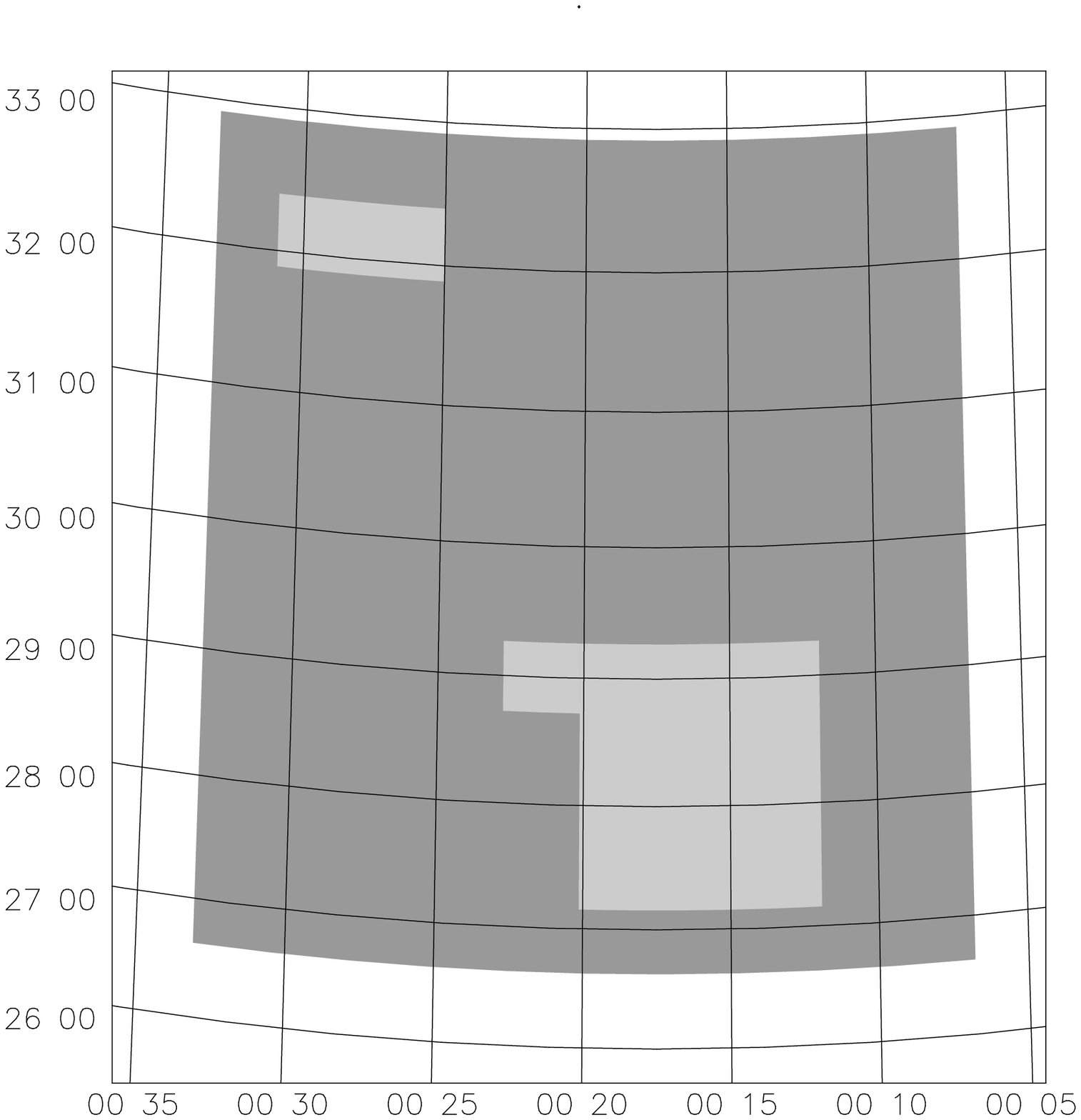,
        angle=270,width=5.8cm,clip=}}\qquad\qquad 
        {\epsfig{file=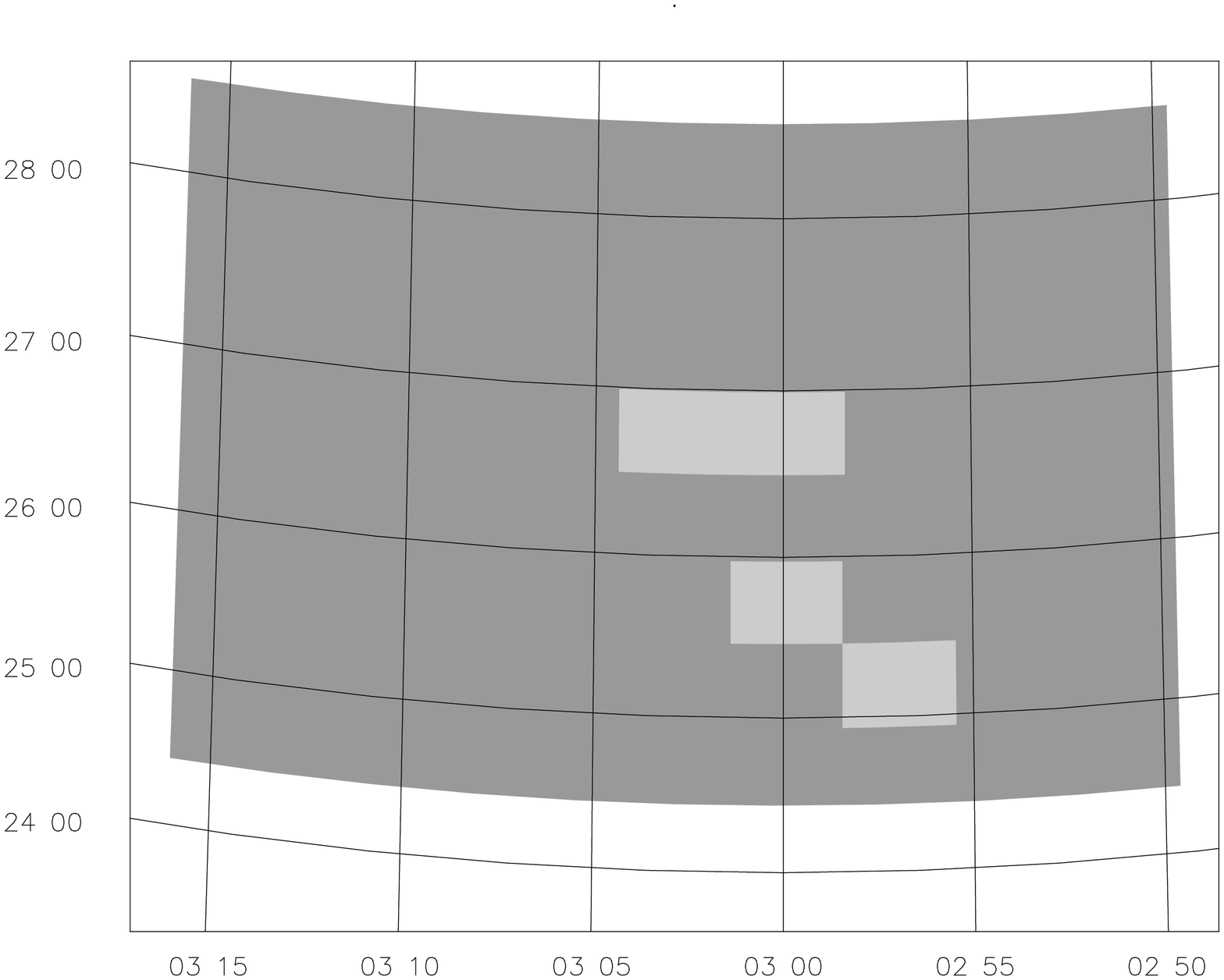,
        angle=270,width=5.8cm,clip=}} 
        {\epsfig{file=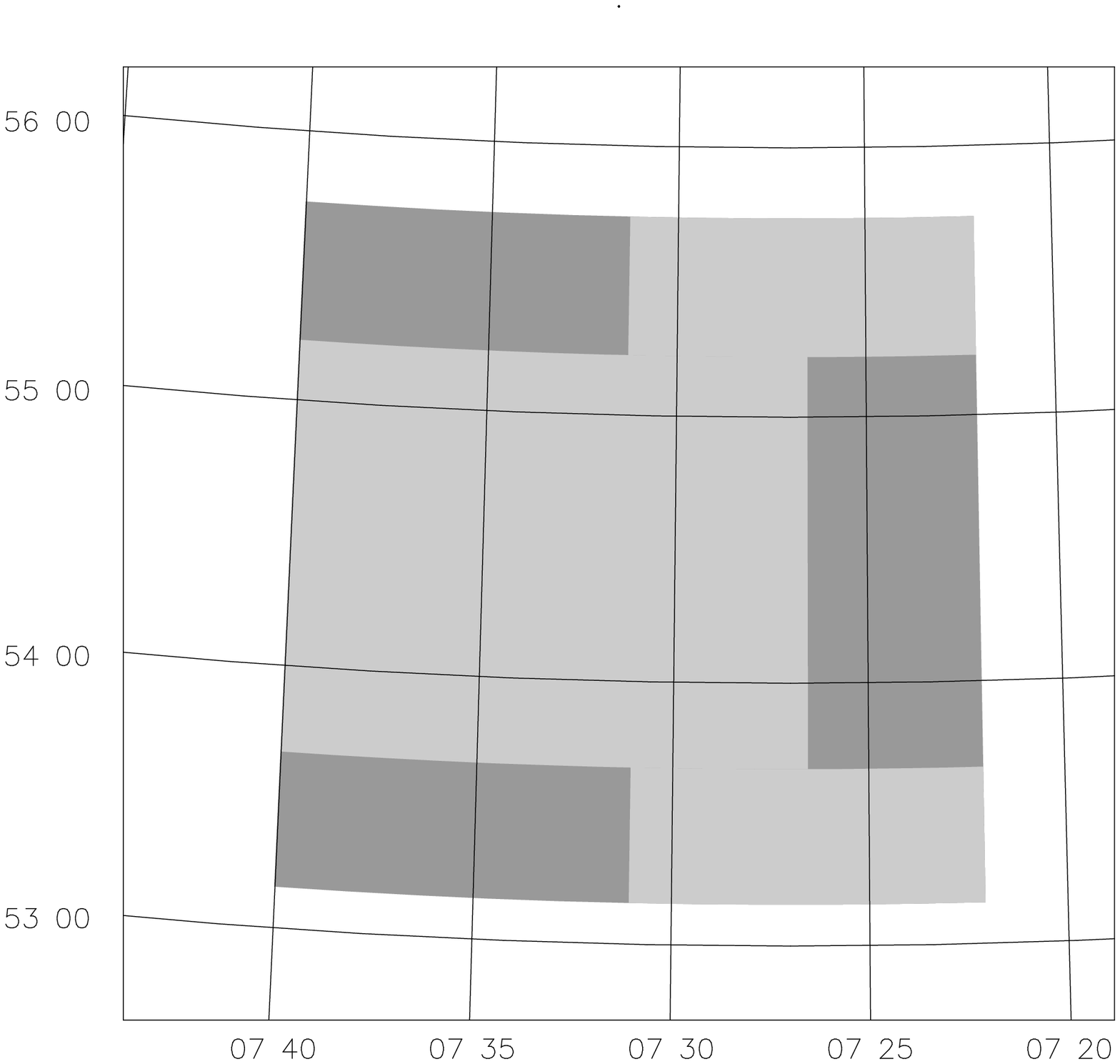,
        angle=270,width=5.8cm,clip=}}\qquad\qquad 
        {\epsfig{file=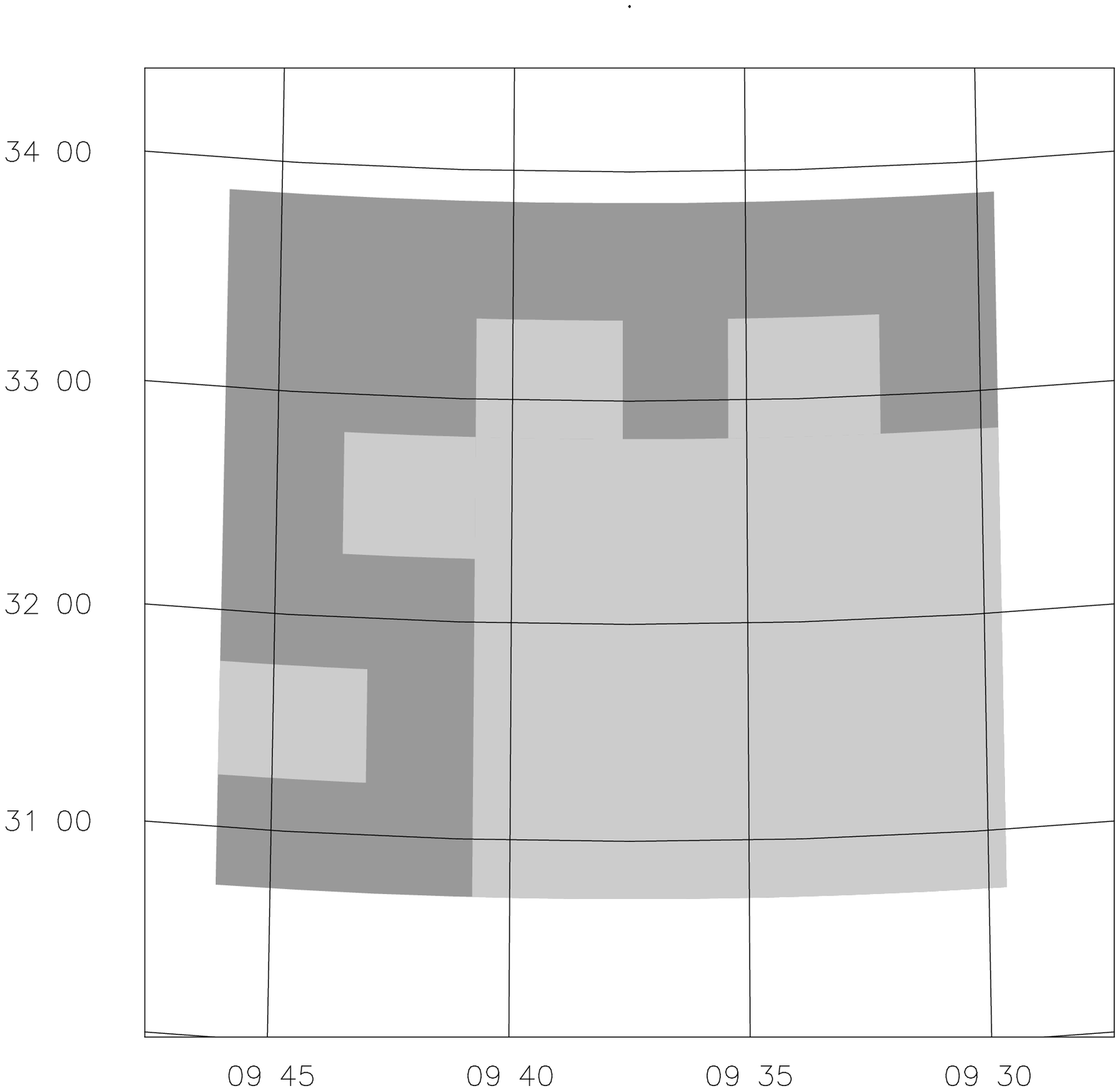,
        angle=270,width=5.8cm,clip=}} 
        {\epsfig{file=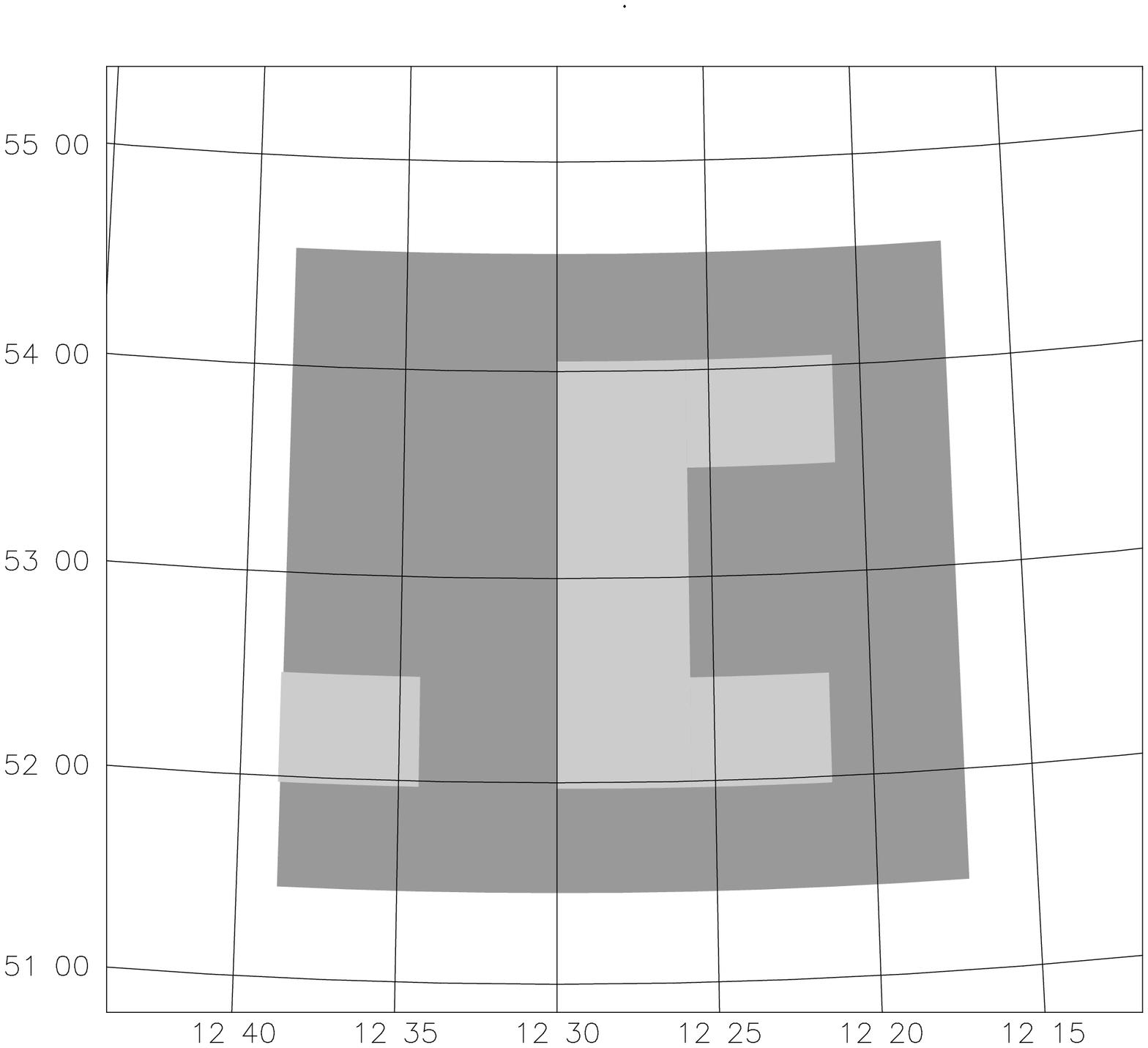,
        angle=270,width=5.8cm,clip=}}\qquad\qquad
        {\epsfig{file=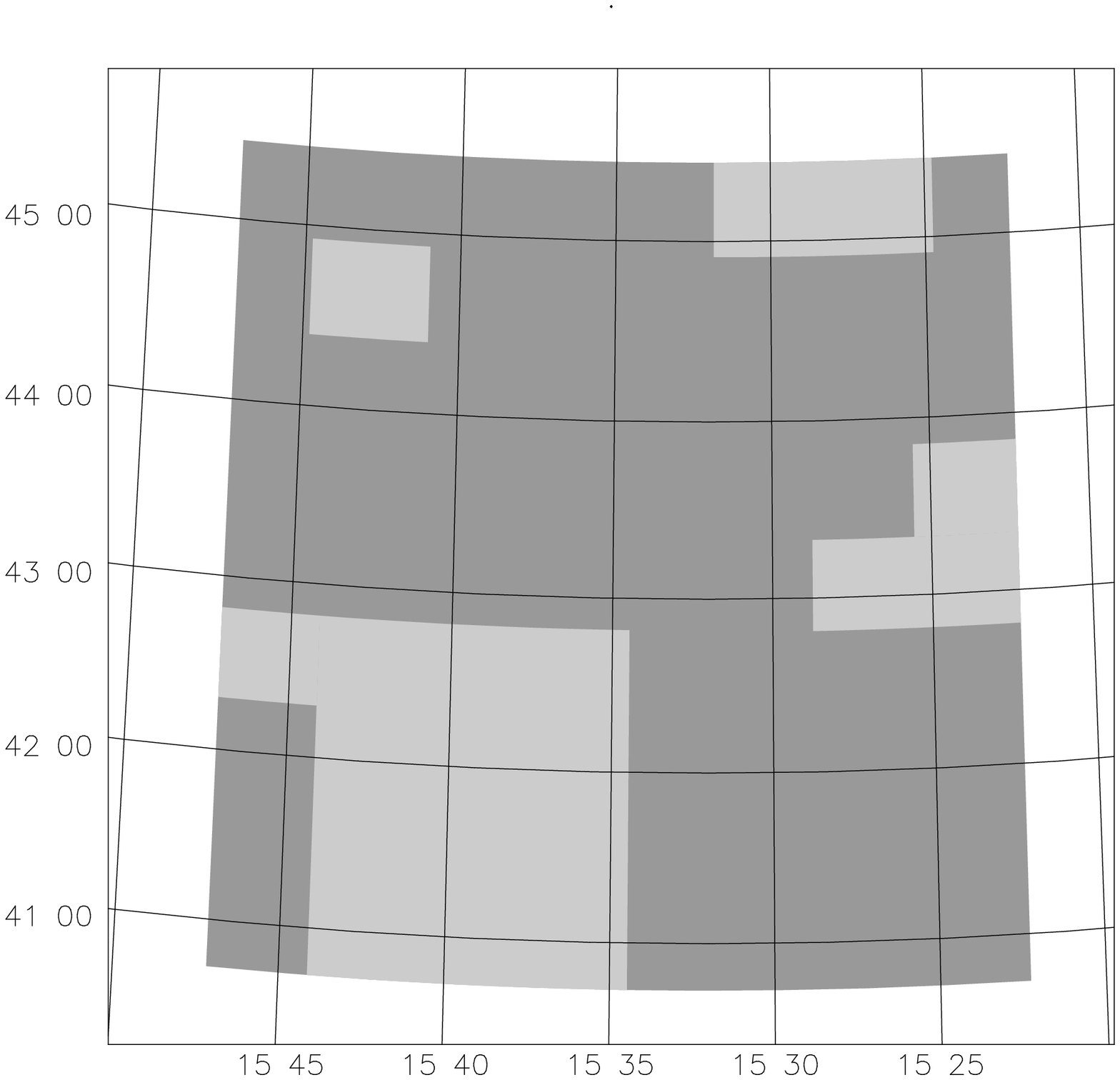,
        angle=270,width=5.8cm,clip=}}
        {\epsfig{file=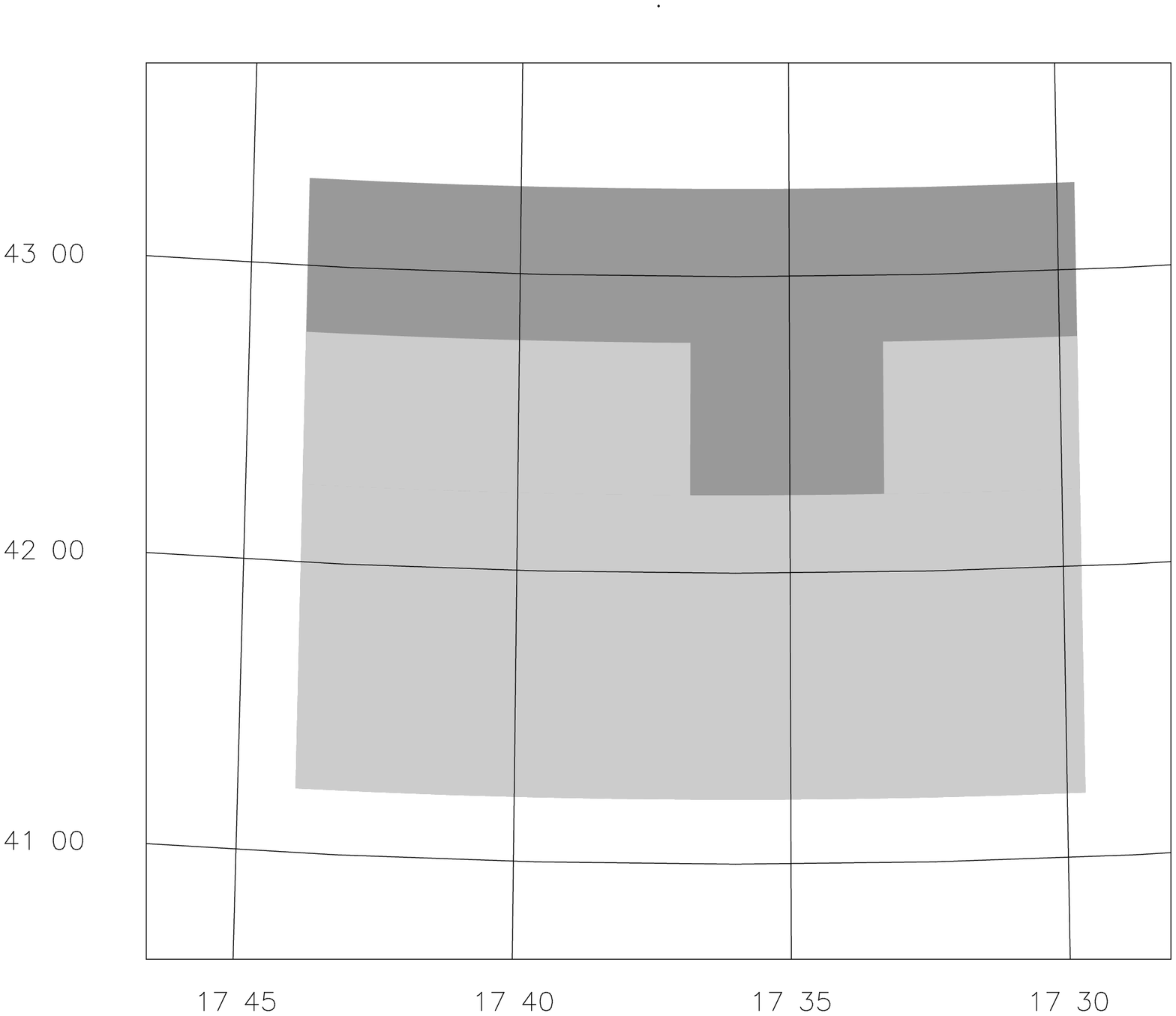,
        angle=270,width=5.8cm,clip=}}
        \caption{The seven fields, centred at 0020+2947, 0303+2629, 0731+5427, 0938+3218, 1228+5301, 1535+4305 and 1735+4215. The total areas are complete to $\approx10$~mJy and the deeper areas (shown in paler grey) to ~$\approx5.5$mJy.}
\label{fig:fields grey}
\end{figure*}

The seven areas of the survey presented here are centred at: 0020+2947, 0303+2629, 0731+5427, 0938+3218, 1228+5301, 1535+4305, 1735+4215 (RA ($^{\rm h}\ ^{\rm m}$) Dec. ($^\circ\ ^\prime$), J2000.0), as indicated in Figure \ref{fig:fields eqproj}. They are situated away from the Galactic plane and are widely spaced in RA, their positions having been determined by the choice of fields for the observations of the VSA. The RA and Dec. ranges are given in Table \ref{table:areastotal}. The total area amounts to 114.7~deg$^{2}$ and within this our source list is complete to $\approx10$~mJy. In each field there are also some much more sensitive areas; this was partly because deeper surveying was required near the centre of the VSA primary beam and partly because on some days there were particularly favourable observing conditions. For the purpose of this paper, in order assemble a complete sample of a useful size, we have selected a number of sub-areas with $\approx5.5$~mJy completeness, which form a total area of 29.1~deg$^{2}$ (Figure~\ref{fig:fields grey}). These are of various shapes and sizes, so in Table~\ref{table:areasdeep} we have described them in terms of small constituent areas bounded by specific RA and Dec. ranges.  Although our completeness limit in these regions is $\approx5.5$~mJy, we have detected many fainter sources, the faintest being only 1 mJy.

\section{Observations and data analysis}

\begin{figure}
        \centerline{\epsfig{file=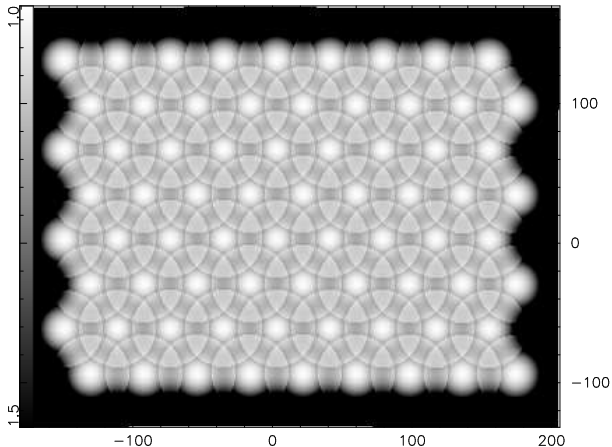,
        angle=270,
        width=8.0cm,clip=}}
        \caption{A typical sensitivity map corresponding to a 9 $\times$ 8 hexagonal array of pointing directions. White indicates high sensitivity. The coordinates are in map pixels, where one pixel represents 8 $\times$ 8 arcsec$^{2}$.}
\label{fig:raster}
\end{figure} 

For our observations we have continued to use a rastering technique similar to that described in Paper I. This is necessitated by the relatively small area within the FWHM of the RT primary beam (0.01~deg$^{2}$) compared with the areas of the VSA fields to be surveyed. As before, we used the RT's E--W array of five aerials to give a resolution of $25 \times 25\mathrm{cosec}\delta$ $\mathrm{arcsec}^2$ and set up a hexagonal pattern of pointing directions with a spacing of 5~arcmin. But, in order to reach the required sensitivity, we now scanned only 72 pointings during a 12h observation, covering an area of sky of only $\sim$~0.3~deg$^{2}$. We also needed to integrate several observations of the same area, sometimes as many as six for the deeper regions, if the weather were unfavourable. Separate so-called `constituent' maps were made from the accumulated data from the individual pointings and, in an improvement on our earlier method, these were then CLEANed before assembling them to form the `raster' map. The method for combining the constituent maps was the same as before and produced a similar variation in sensitivity (of $\sim$~20~\%) over the corresponding raster map, as illustrated in Figure~\ref{fig:raster}. We made an estimate of the mean noise, $\sigma$, over each raster map and continued to integrate data until we had reached the required sensitivity. Over all the areas we reached $\sigma\leq2$ mJy and in the deeper areas $\sigma\leq1$ mJy, though there was considerable variation within these.

The source extraction was also implemented much as before. Possible sources were identified as peaks from interpolation of the raster map and then each of these was followed up with a short pointed observation of 10--15 minutes, either to establish a reliable flux density or to eliminate it as a false detection. The only change in our method was that, while we continued to take $3\sigma$ as the pixel cut-off on the raster map, we now took $4.5\sigma$, rather than $5.0\sigma$, as the cut-off for the interpolated peak, in order to find as many of the weaker sources as possible.     

\section{Completeness}

\begin{figure}
        \centerline{\epsfig{file=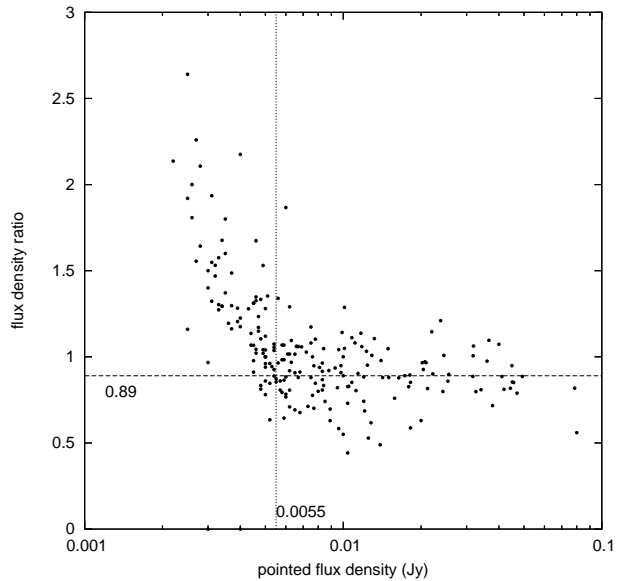,
        angle=270,
        width=8.0cm,clip=}}
        \caption{Plot of the ratio $(S_{\mathrm{r}}/S_{\mathrm{p}})$ of raster flux density to pointed flux density versus $S_{\mathrm{p}}$, for the deeper areas. The estimated completeness limit is $\approx5.5$mJy. (Where the ratio is shown as zero, the corresponding source lies in an area of a raster map which is hard to interpret. A pointed observation has been made but there is no definitive raster flux density.)}
\label{fig:fratio}
\end{figure}

Since the original candidate sources were found from raster maps with a range of noise levels, assessment of the completeness limits of our source lists is not straightforward. Although we can be sure that every source in our catalogue is  genuine, since it has been followed up with a pointed observation, it is difficult to estimate the number of undetected sources at any flux density level. We have approached the problem in the following way.

Taking the deeper areas first, we have plotted the ratio, $(S_{\mathrm{r}}/S_{\mathrm{p}})$, of the raster flux density to the pointed flux density versus $S_{\mathrm{p}}$, as in Figure \ref{fig:fratio}. We find that, for the brighter sources, in the range $\geq10$~mJy where we can assume completeness, there is a scatter about a median ratio of $\approx0.89$.  This off-set of $\approx10$~per~cent in the median value is probably attributable to pointing problems; in particular, there was a necessary compromise between keeping as many of the observed data samples as possible and discarding those with excessive pointing errors. For the weaker sources the ratio rises progressively with decreasing flux density because at these levels the peaks detected on the raster maps are preferentially those boosted by a positive local noise contribution. In order to estimate the completeness limit, we have investigated the change in the relative numbers of sources above and below the median line as the pointed flux density decreases from 10~mJy. We find that there is no significant difference in the numbers down to 5.5~mJy but below this level there are significantly more sources with ratios above 0.89 than below. We therefore assume we have completeness to approximately 5.5~mJy, which means there are 135 sources above the limit and 81 below.

For the total areas, we apply similar arguments using higher flux densities and conclude we have a completeness limit of approximately 10~mJy, giving us 307 sources above the limit and 336 below. 

We should also explain that within the areas presented in this paper there is a shortfall in the number of sources above 100~mJy as compared with the number expected from the source count in Paper I (see next section). This is because these VSA fields were deliberately selected to contain as few bright sources as possible.

\section{Source counts}

\begin{table}
\centering
\caption{The data used in the combined count}
\begin{tabular}{ccccccc}
\hline
          bin start    &&    bin end    &&    number    &&    area \\
          $S$/Jy       &&    $S$/Jy     &&    $N$       &&    sr   \\
\hline
          0.0055     &&       0.0063     &&       23      &&       0.008856 \\
          0.0063     &&       0.0078     &&       21      &&       0.008856 \\
          0.0078     &&       0.0100     &&       21      &&       0.008856 \\ 
          0.0100     &&       0.0110     &&       38      &&       0.034939 \\
          0.0110     &&       0.0129     &&       43      &&       0.034939 \\
          0.0129     &&       0.0152     &&       40      &&       0.034939 \\
          0.0152     &&       0.0198     &&       40      &&       0.034939 \\
          0.0198     &&       0.0250     &&       39      &&       0.034939 \\
          0.025      &&       0.030      &&       80      &&       0.15841  \\
          0.030      &&       0.040      &&      104      &&       0.15841  \\
          0.040      &&       0.060      &&      103      &&       0.15841  \\
          0.060      &&       0.100      &&       94      &&       0.15841  \\
          0.100      &&       0.200      &&       49      &&       0.15841  \\
          0.200      &&       0.500      &&       27      &&       0.15841  \\
          0.500      &&       1.000      &&        8      &&       0.15841  \\
\hline
\end{tabular}
\label{table:count_data}
\end{table}

\begin{figure}
        \centerline{\epsfig{file=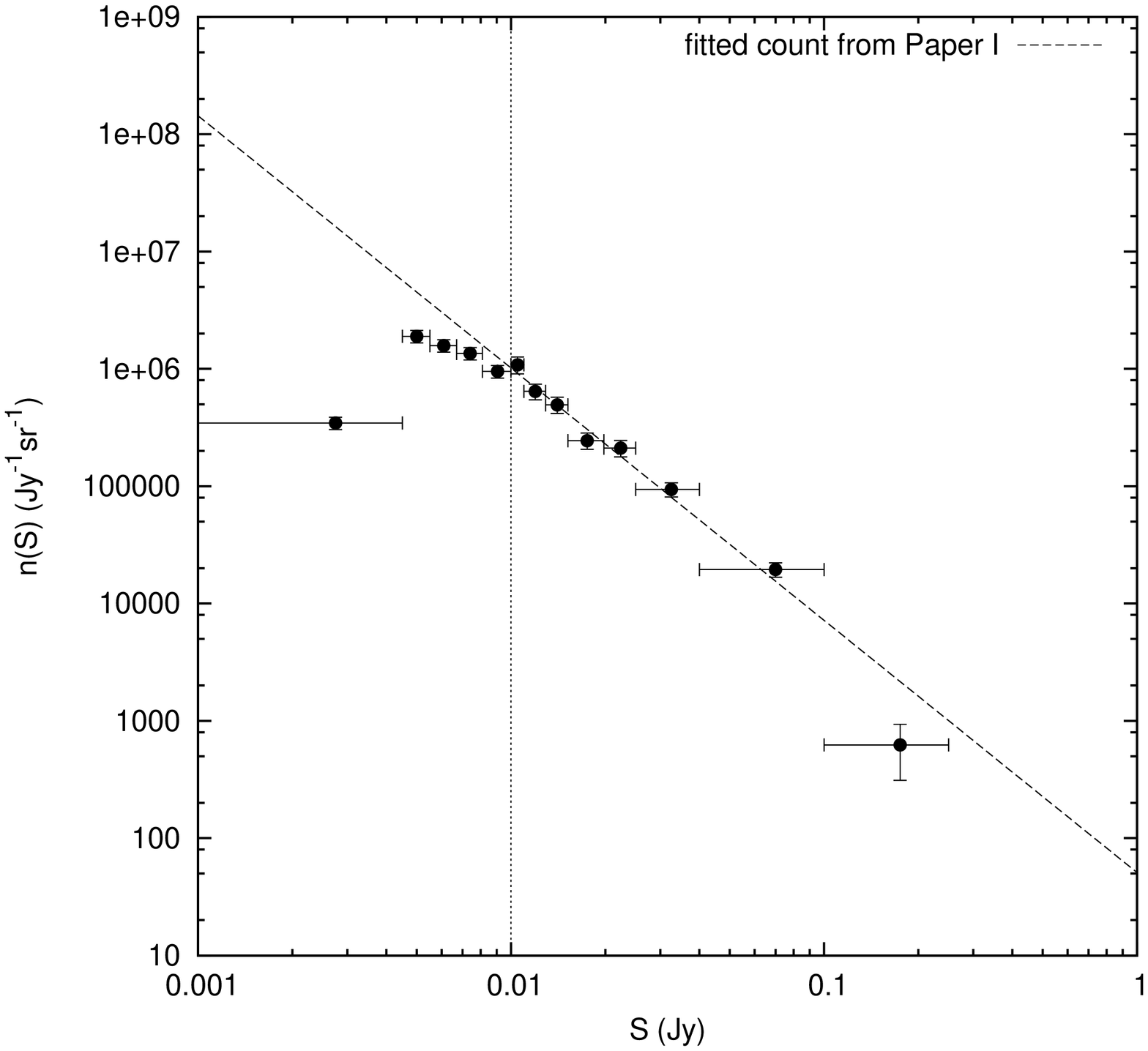,
        angle=270,
        width=8.0cm,clip=}}
\caption{The source count for the sources in the total areas, showing the lower completeness limit of $\approx10$mJy and the shortfall above 100~mJy}
\label{fig:count_total}
\end{figure}

\begin{figure}
        \centerline{\epsfig{file=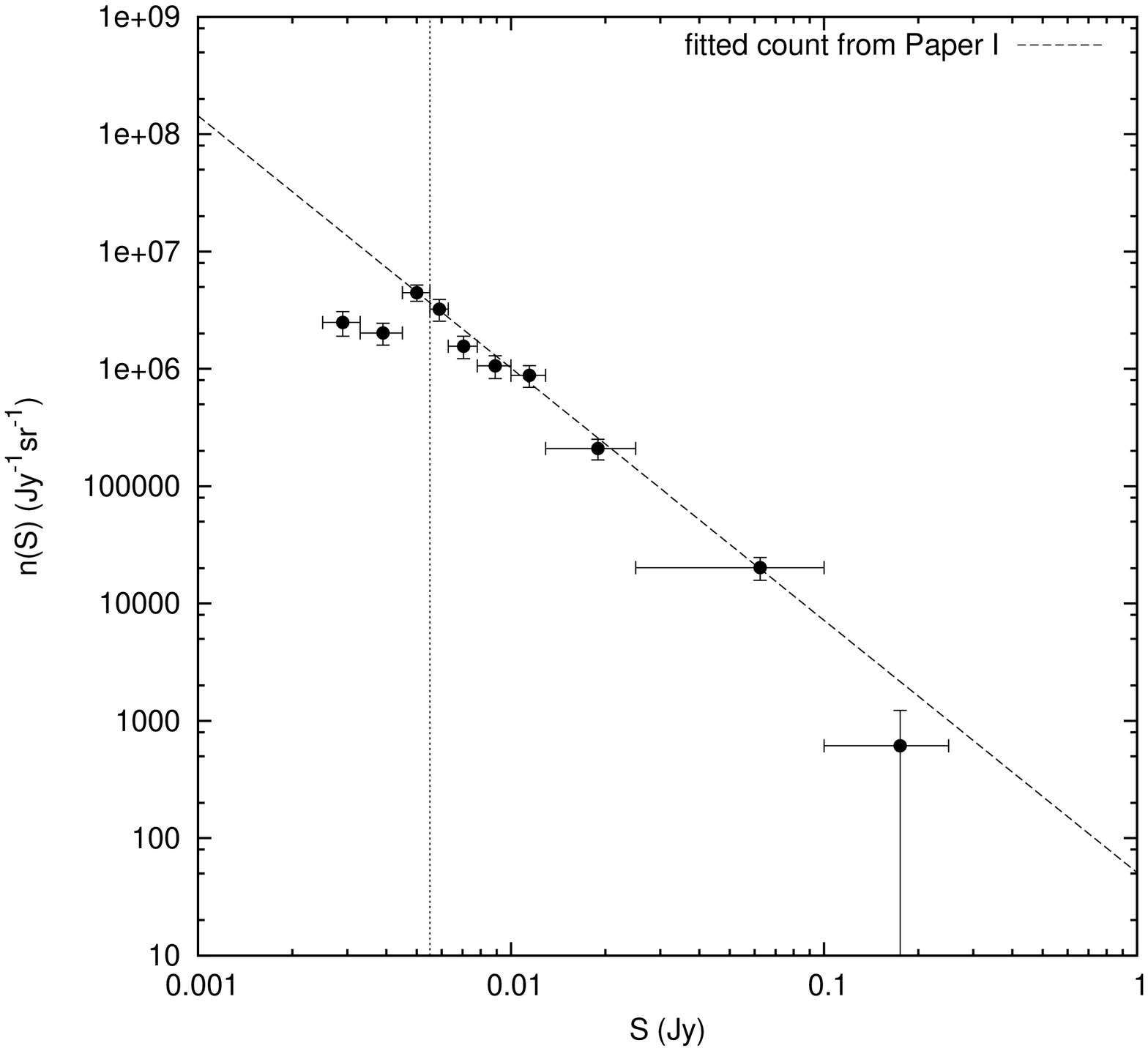,
        angle=270,
        width=8.0cm,clip=}}
\caption{The source count for the sources in the deeper areas, showing the lower completeness limit of $\approx5.5$mJy and the shortfall above 100~mJy}
\label{fig:deep_count}
\end{figure}

\begin{figure}
        {\epsfig{file=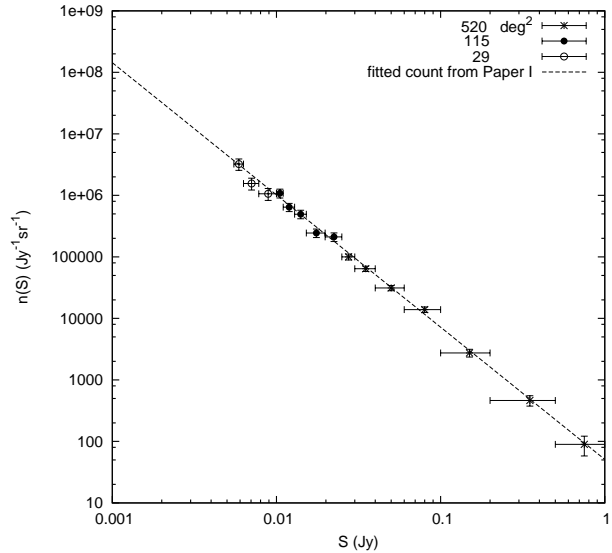,
        angle=270,width=8.0cm,clip=}}
        \caption{Differential  count extended to 5.5~mJy}
\label{fig:comb_count}
\end{figure}

\begin{figure}
        {\epsfig{file=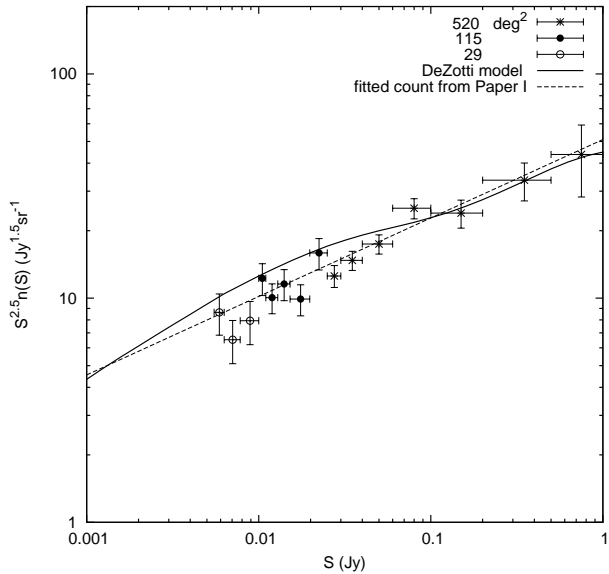,
        angle=270,width=8.0cm,clip=}}
        \caption{Normalised count compared with the de Zotti model}
\label{fig:norm_count}
\end{figure}

The source count for the sources in the total areas is shown in Figure \ref{fig:count_total} which indicates the lower completeness limit of 10~mJy and also the shortfall above 100~mJy mentioned in the previous section. Figure \ref{fig:deep_count} is a similar plot for the sub-set of these sources which lie in the deeper areas.

Using these counts, we are now able to extend the original 9C count to a lower flux density level. We have combined data from Paper I in the range $S\geq25$~mJy with data from this paper in the range $10 \leq S < 25$~mJy from the total areas and $5.5 \leq S < 10$~mJy from the deeper areas (see Table \ref{table:count_data}). As explained in the earlier paper, the effect of the random errors in the flux densities is negligible compared with the Poisson errors on $N$, but we have applied a correction for the bin widths in the same way as before. At the completeness limit of 5.5~mJy the survey continues to be limited by noise rather than confusion; we estimate that above this level there are typically 2000 synthesized beam areas per source. 

Figure~\ref{fig:comb_count} shows the differential count $n(S)$ and we see that, down to 5.5~mJy, there is no evidence for any significant change from the fitted count $n(S)$ calculated in Paper I: i.e.
\[
n(S) \equiv \frac{{\rm d}N}{{\rm d}S} \approx 51 \left( \frac{S}{\rm Jy} \right)
^{-2.15}
\, {\rm Jy}^{-1}{\rm sr}^{-1}
\] 
In Figure \ref{fig:norm_count} we compare our normalized count $S^{2.5}n(S)$ with that derived from the model of de Zotti et al. (2005) and note that our count appears to lie somewhat below their prediction in the lower flux density range.

These two figures effectively update the plots in Figures 6 and 7 of Paper I. We should mention here that we found retrospectively that the positions of two points in each of those figures are in error. These are the two open circles corresponding to the higher flux bins of the deeper survey. In fact, since they lie in the flux density range of the main survey, they do not affect any of the arguments or conclusions in that paper and can simply be ignored.   

\section{Correlation with NVSS at 1.4~GHz}

In this section we present the results of our investigations but defer any discussion of their implications to Section~8.

\subsection{Matching the catalogues}

\begin{figure}
        \centerline{\epsfig{file=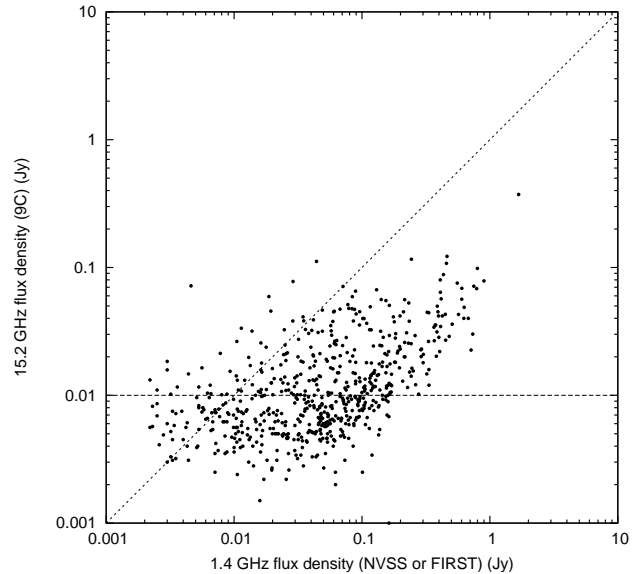,
        angle=270,
        width=11.0cm,clip=}}
        \caption{Plot of RT 15.2~GHz pointed flux densities versus NVSS (or FIRST) 1.4~GHz flux densities, with a line indicating zero spectral index.  The horizontal line is the estimated completeness limit.}
\label{fig:9C-NVSS}
\end{figure}

We have attempted to match the 643 sources in our catalogue with the catalogue from the NVSS 1.4-GHz survey, which has a resolution of 45~arcsec (not dissimilar to our own) and a completeness of 50~percent at $\approx2.5$~mJy rising rapidly to 99~percent at 3.4~mJy. In the first instance, we searched for a counterpart within 40~arcsec of each of the 9C sources and then in every case inspected the corresponding NVSS contour plot, paying particular attention to sources which had multiple matches within 40~arcsec and to those where there was no match within 20~arcsec.   For about 90~percent of the sources the matching was straightforward; the remainder, however, required further attention in order to make the appropriate correlation of flux densities. For example, some binary sources had a single entry in the 9C catalogue, but appeared separately in the NVSS catalogue and, conversely, there were some single NVSS sources which appeared as separate components in 9C. Where possible, we made use of the deeper FIRST 1.4-GHz survey (Becker, White \& Helfand 1995), with its resolution of 5~arcsec, to assist in resolving ambiguities.

Our correlation is illustrated in Figure \ref{fig:9C-NVSS}. In all, out of the 643 sources in our list, there were 21 with no counterpart in the NVSS catalogue, though two of these did appear in FIRST. Also, a number of them were found to be coincident with one or two contours on the NVSS plots but clearly these were not significant enough detections to justify their inclusion as sources in the NVSS catalogue. (See sub-section 6.3 for further details.)

\subsection{Spectral index distributions: samples selected at 15.2~GHz}

\begin{figure}
        \centerline{\epsfig{file=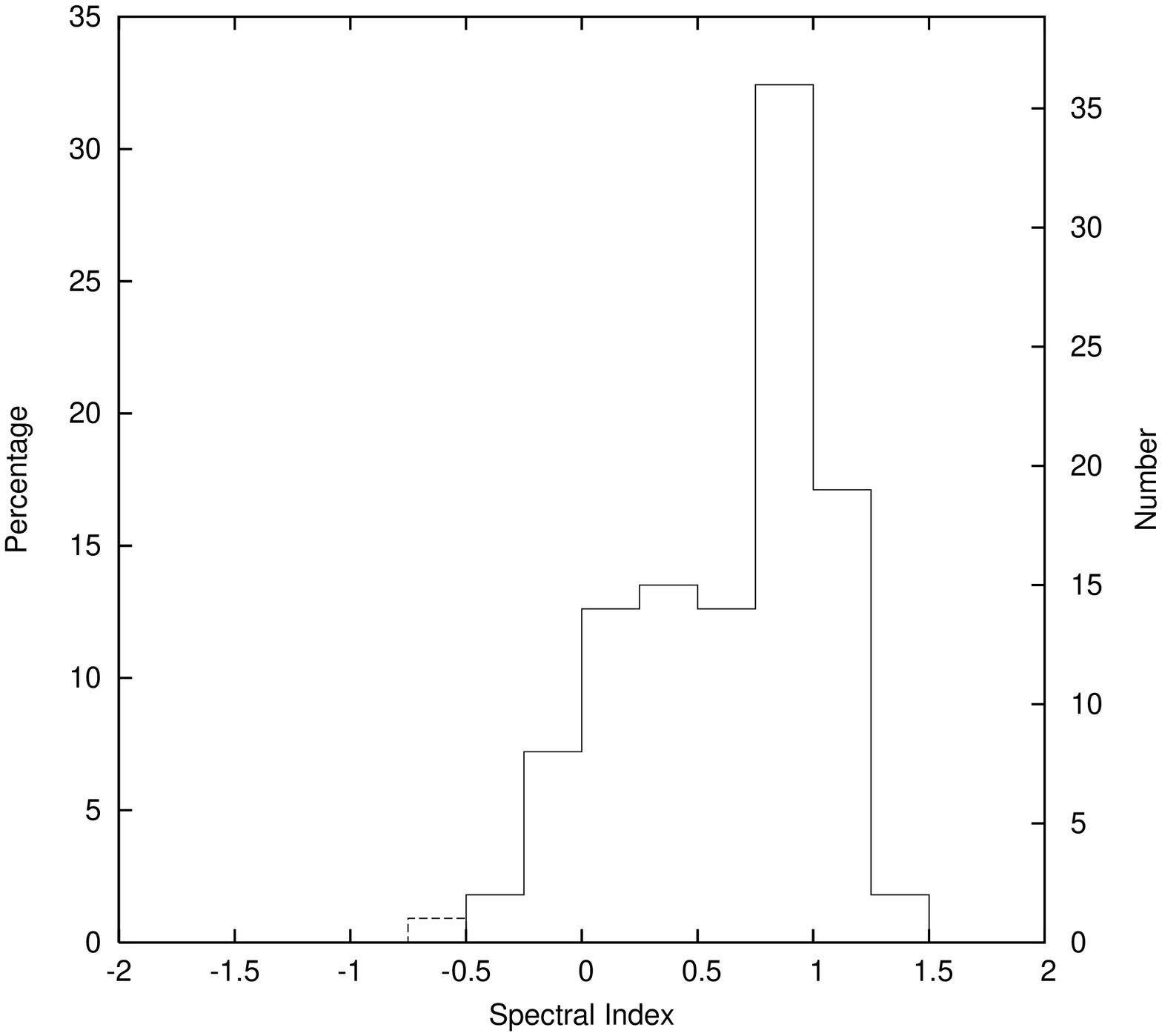,
        angle=270,
        width=7.5cm,clip=}}
        \centerline{\epsfig{file=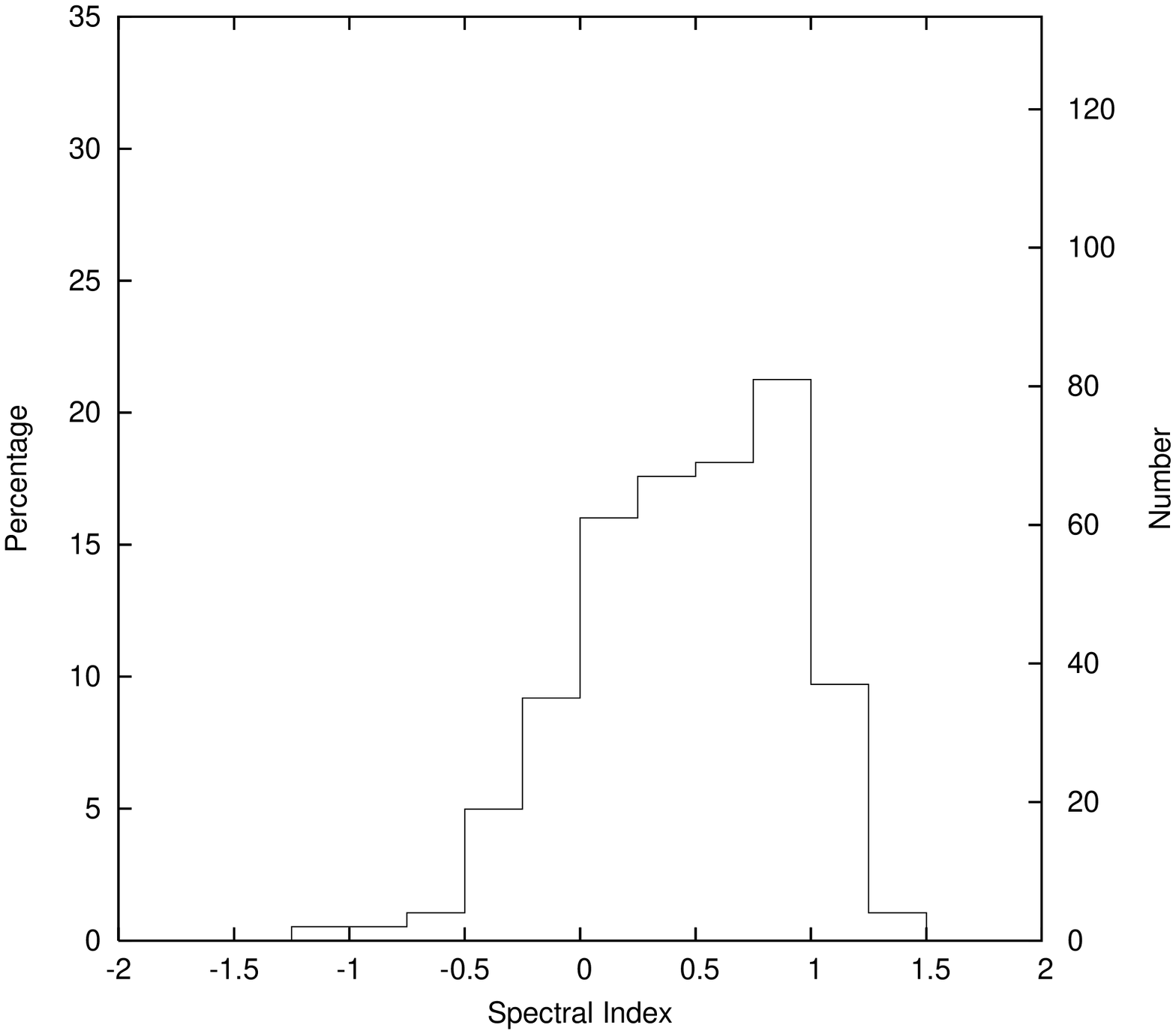,
        angle=270,
        width=7.5cm,clip=}}
        \centerline{\epsfig{file=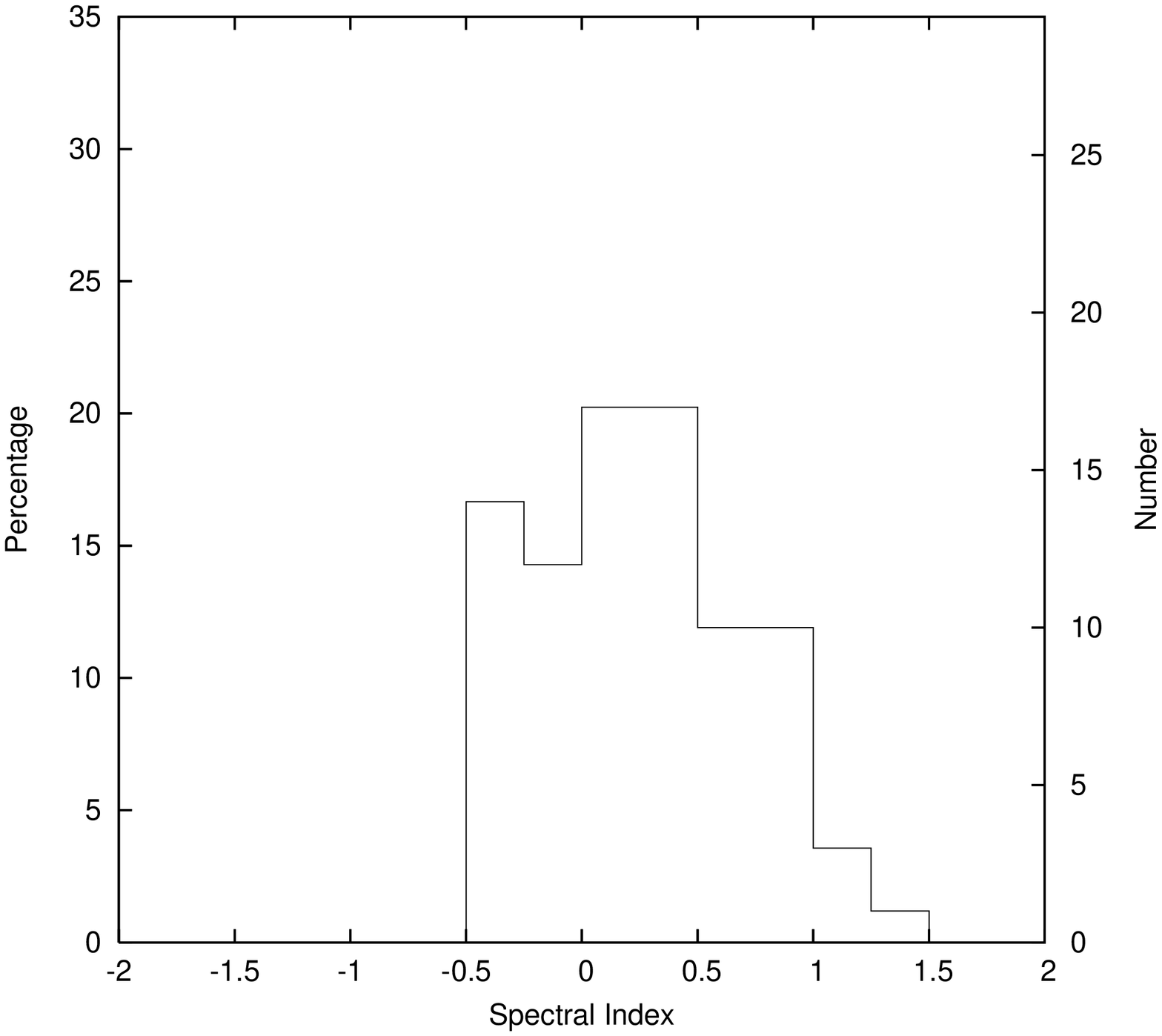,
        angle=270,
        width=7.5cm,clip=}}
        \caption{The $\alpha_{1.4}^{15.2}$ spectral index distributions for samples selected at 15.2~GHz in three flux density ranges. From the top: $5.5 \leq S < 25$ mJy, $25 \leq S < 100$ mJy and $S\geq100$ mJy, with median values of 0.79, 0.51 and 0.23 respectively. We define $\alpha$ by $S\propto \nu^{-\alpha}$. (The dotted box in the top plot indicates a limiting value of $\alpha$ for a source not in NVSS)}
\label{fig:alphas_1}
\end{figure}

\begin{table}
 \caption{Some statistics for the the spectral index distributions $\alpha_{1.4}^{15.2}$}
 \begin{tabular}{@{}c c c c c }
 \hline
 15.2-GHz flux&Number&Number of&Number of&$\tilde{\alpha}_{1.4}^{15.2}$\\
 density&of&sources&sources&\\
 range (mJy)&sources&with&with&\\
 &&$\alpha_{1.4}^{15.2} \leq 0$&$\alpha_{1.4}^{15.2} \leq 0.5$&\\
 \hline
 $5.5 \leq S < 25$&111&11 (10~\%)&40 (36~\%)&0.79\\
 $25 \leq S < 100$&381&62 (16~\%)&190 (50~\%)&0.51\\
 $S \geq 100$&84&26 (31~\%)&60 (71~\%)&0.23\\
 \hline
 \end{tabular}
\label{table:alphas_1}
\end{table}

\begin{figure}
        \centerline{\epsfig{file=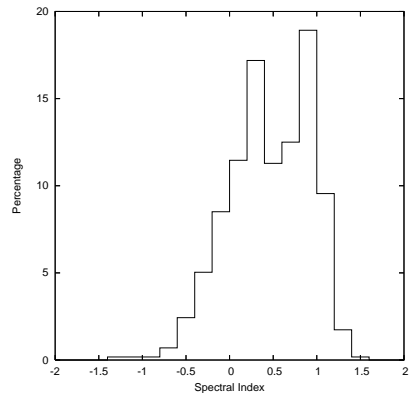,
        angle=270,
        width=7.5cm,clip=}}
        \caption{The $\alpha_{1.4}^{15.2}$ spectral index distributions in Figure~\ref{fig:alphas_1} combined into one distribution. $S\propto \nu^{-\alpha}$ }
\label{fig:alphas_2}
\end{figure}

\begin{table*}
\begin{minipage}{130mm}
 \caption{Results from a Monte Carlo simulation, of one million realisations, to assess the significance of the shift in median spectral index with flux density.}
 \begin{tabular}{@{}c c c c c c c }
 \hline
 15-GHz flux&Number&Min.&Max.&Mean&Number of&Probability of\\ 
 density&of&$\tilde{\alpha}_{1.4}^{15.2}$&$\tilde{\alpha}_{1.4}^{15.2}$&$\tilde{\alpha}_{1.4}^{15.2}$&realisations&drawing median\\
 range (mJy)&simulated&&&&with $\tilde{\alpha}_{1.4}^{15.2}$&more extreme\\
 &sources&&&&&than observed\\
 \hline
 $5.5 \leq S < 25$&111&0.166&0.817&0.477&$> 0.79$ is 14&$2.8\times10^{-5}$ \\
 $25 \leq S < 100$&381&0.311&0.685&0.474&$> 0.51$ is 212833&0.426 \\
 $S \geq 100$&84&0.081&0.879&0.478&$< 0.23$ is 161&$3.2\times10^{-4}$ \\
 \hline
 \end{tabular}
\label{table:montecarlos}
\end{minipage}
\end{table*}

A histogram of the spectral index distribution $\alpha_{1.4}^{15.2}$ for the 111 sources with $5.5 \leq S < 25$~mJy and lying in our deep areas is shown in Figure \ref{fig:alphas_1}.  We define $\alpha$ by $S\propto \nu^{-\alpha}$.  The figure also shows the spectral index distributions, over the same range of frequencies, for sources with $25 \leq S < 100$~mJy and $S \geq 100$~mJy; these distributions contain 381 and 84 sources, respectively, and were originally included in Paper I.  They are derived from an area of 520~deg$^{2}$ complete to $\approx25$~mJy. In Table~\ref{table:alphas_1} we have summarised some important features of the distributions; it is clear from this table and Figure~\ref{fig:alphas_1} that, as one moves to lower flux densities, the proportion of sources with flat and rising spectra decreases.  

In order to assess the significance of the shift in median spectral index with flux density, we have combined the spectral indices of the 576 sources to form a single spectral index distribution, which we have plotted as a histogram in Figure~\ref{fig:alphas_2}.  We assume that there is an underlying spectral index distribution which has no dependence on flux density and that the combined data represent our best approximation to this distribution.  For each of our flux density bins, we calculate the probability of drawing the observed number of sources from the combined distribution and finding a median spectral index as extreme as those observed, using a Monte Carlo simulation of $10^{6}$ realisations. For example, for the lowest flux density bin, for each realisation, we draw, at random, 111 values from the combined spectral index distribution and find the median spectral index.  We find 14 realisations, out of $10^{6}$, in which the this value is greater than or equal to the observed value of 0.79 and conclude that the probability of drawing 111 sources from the combined distribution and observing a median as or more extreme than 0.79 is $2.8\times10^{-5}$.

Table~\ref{table:montecarlos} provides a summary of the results from the simulations. It shows that the median spectral index observed for the middle flux density bin is not significantly different from that of the combined distribution of Figure~\ref{fig:alphas_2}.  However, the probabilities of having obtained median spectral indices more extreme than observed, for the two outer flux density bins, by drawing sources randomly from the combined distribution are very small.  Consequently, we can reject our null hypothesis, that the underlying spectral index distribution has no dependence on flux density, with a high level of confidence.  In other words, the shift in the median spectral index with flux density is significant.

\subsection{Spectral index distributions: samples selected from NVSS at 1.4~GHz}
\begin{figure}
        \centerline{\epsfig{file=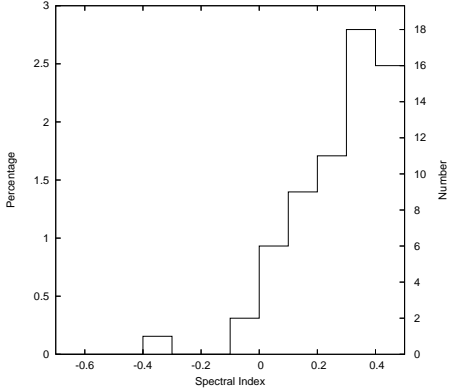,
        angle=270,
        width=7.5cm,clip=}}
        \caption{Histogram of the spectral index distribution $\alpha_{1.4}^{15.2}$ for a sample selected from NVSS with $S_{1.4} \geq 32.95$~mJy, showing the percentage of sources with flat and rising spectra. $S\propto \nu^{-\alpha}$ }
\label{fig:alphas_3}
\end{figure}

\begin{table*}
\begin{minipage}{130mm}
 \caption{Proportions of rising, and flat and rising sources in the NVSS catalogue.}
 \begin{tabular}{@{}c c c c c }
 \hline
 Flux density&Number of&9C Area&Select&Number of\\
 cut at&NVSS&&sources&sources\\
 1.4~GHz (mJy)&sources&&with&selected\\
 \hline
 32.95&644&110~deg$^{2}$ complete to 10~mJy&$\alpha_{1.4}^{15.2} \leq 0.5$&63 (9.8~\%)\\
 10.0&1891&110~deg$^{2}$ complete to 10~mJy&$\alpha_{1.4}^{15.2} \leq 0$&20  (1.1~\%)\\
 18.12&295&29~deg$^{2}$ complete to 5.5~mJy&$\alpha_{1.4}^{15.2} \leq 0.5$&27 (9.2~\%)\\
 5.5&770&29~deg$^{2}$ complete to 5.5~mJy&$\alpha_{1.4}^{15.2} \leq 0$&11 (1.4~\%)\\
 \hline
 \end{tabular}
\label{table:flat_rising}
\end{minipage}
\end{table*}

It is not possible to calculate a full spectral index distribution for a sample of NVSS sources complete to a level of 3.4~mJy, since many of the steep spectrum sources will not be detectable in our survey. We can, however, investigate the proportion of flat and rising spectrum sources in the NVSS catalogue for different cut-offs in 1.4~GHz flux density, where we use `flat and rising' to refer to sources with $\alpha_{1.4}^{15.2} \leq 0.5$.   

We have selected all sources from the NVSS catalogue, lying in the areas complete to 10~mJy at 15.2~GHz, with $S_{1.4} \geq 32.95$~mJy; in total, 644 sources are selected.  These selection criteria ensure that any sources with $\alpha_{1.4}^{15.2} \leq 0.5$ will have $S_{15.2} \geq 10$~mJy, our completeness limit at 15.2~GHz.  Of the 644 sources selected, 63 are found to have $\alpha_{1.4}^{15.2} \leq 0.5$.  In other words, we find that 9.8~per~cent of sources with $S_{1.4} \geq 32.95$~mJy have $\alpha_{1.4}^{15.2} \leq 0.5$.  We have plotted this flat and rising part of the spectral index distribution $\alpha_{1.4}^{15.2}$ at 1.4~GHz as a histogram in Figure~\ref{fig:alphas_3}. We have also taken a lower flux density cut at 1.4~GHz of 10~mJy (again in our total areas), which enables us to investigate sources with $\alpha_{1.4}^{15.2} \leq 0$ for a somewhat deeper sample.  

Similarly, we have taken appropriate NVSS flux density cuts in the deeper areas, complete to 5.5~mJy at 15.2~GHz, to enable us to investigate the rising, and flat and rising side of the distribution at lower flux densities.  Our results are summarised in Table~\ref{table:flat_rising}.  

\subsection{Sources not in NVSS} 

\begin{table}
 \caption{9C sources without corresponding matches in NVSS.  Flux densities for those sources appearing in the FIRST survey are provided, along with the spectral index based upon the FIRST flux density.  For sources not appearing in FIRST, an upper limit to the spectral index is provided by assuming a flux density of 3.4 mJy (the estimated 99 per cent completeness limit of NVSS) at 1.4 GHz.}
 \begin{tabular}{@{}c c c c c }
 \hline
 RA (J2000)&Dec. (J2000)&15.2-GHz&1.4-GHz&$\alpha$\\
 &&flux&flux&\\
 &&density&density&\\
 &&(mJy)&(mJy)&\\
 \hline
 17:40:17.1&+42:14:31&14.0&&-0.59\\
 02:53:29.4&+27:57:45&13.0&&-0.56\\
 00:27:28.0&+28:08:10&10.9&&-0.49\\
 00:14:19.0&+31:33:45&10.2&&-0.46\\
 15:37:42.1&+43:49:12&10.2&0.87&-1.03\\
 12:35:22.9&+53:11:29&6.2&&-0.25\\
 00:11:51.7&+27:57:19&6.0&&-0.24\\
 09:44:27.2&+33:03:22&6.0&&-0.24\\
 09:37:22.6&+32:01:04&5.3&&-0.19\\
 09:34:35.9&+30:44:05&5.0&&-0.16\\
 09:30:36.0&+30:48:54&4.5&&-0.12\\
 17:33:11.1&+41:21:05&4.5&&-0.12\\
 00:27:10.3&+26:55:42&4.4&&-0.11\\
 15:32:04.6&+42:51:52&4.1&2.56&-0.20\\     
 03:14:20.8&+26:26:27&4.0&&-0.07\\
 15:38:47.8&+43:51:18&3.5&&-0.01\\
 17:42:52.2&+43:09:51&3.2&&0.03\\
 00:15:58.0&+27:18:54&3.1&&0.04\\
 00:31:12.7&+27:12:58&3.0&&0.05\\
 07:27:18.0&+55:10:45&3.0&&0.05\\
 15:40:52.8&+44:16:22&2.6&&0.11\\
 \hline
 \end{tabular}
\label{table:not_in_nvss}
\end{table}

\begin{table}
 \caption{9C sources with no NVSS counterpart as a function of flux density.}
 \begin{tabular}{@{}c c c }
 \hline
 Flux&Number&Number of\\
 density&of 9C&sources with no\\
 range (mJy)&sources&NVSS counterpart\\
 \hline
 $S < 4.2$&62&8 (12.9~\%)\\
 $4.2 \leq S < 10$&281&8 (2.8~\%)\\
 $10 \leq S < 15$&117&5 (4.3~\%)\\
 \hline
 \end{tabular}
\label{table:not_in_nvss_stats}
\end{table}

Table~\ref{table:not_in_nvss} lists the 21 sources in our total source list which do not appear in the NVSS catalogue. Only one of these lies in the deeper areas (see the dotted box in the top plot of Figure~\ref{fig:alphas_1}) and the others are all in the total areas. By dividing these 21 sources into three flux-density bins, as shown in Table~\ref{table:not_in_nvss_stats}, we can gain an indication of the proportion of 9C sources, with no counterparts in NVSS, as a function of flux density. We are complete only in the highest bin ($10 \leq S < 15$~mJy), for which we find 4.3~per~cent of sources with no NVSS counterpart.  Assuming that the proportion of sources with no counterpart increases with decreasing flux density, our estimates of the proportions of `missing' sources are likely to be under-estimates for the two lower flux density bins.  This is because these bins are incomplete; sources with higher fluxes are preferentially detected in these incomplete bins, yet it is the brighter sources that are more likely to have an NVSS counterpart.

We should note that several of the sources have been observed at other frequencies.  For example, a number are close to sources in the 2MASS (Skrutskie et al. 2006) and \textit{ROSAT} all-sky survey (Voges et al. 1999) catalogues.  However, here we are concerned with the number of sources at 15~GHz that would be missed by reliance on the NVSS catalogue.

\section{The source catalogue}

\begin{table*}
\caption{A section from the source catalogue. The letter `D' means that the source lies in one of the deeper areas listed in Table \ref{table:areasdeep}. In the comment column `b' denotes a binary source and `e' an extended source.}
\begin{tabular}{lcccccccc}
\hline
Source name  &&&    RA J2000 &  Dec. J2000     &&   Flux density &  Date & Comment \\
&&&$^{\rm h}\ ^{\rm m}\ ^{\rm s}$&$^\circ\ ^\prime\ ^{\prime\prime}$&& (Jy) &yymmdd \\
\hline
9CJ0015+3123 &   &&     00:15:47.5   &     31:23:35 &&   0.0127 &   020505 &    e \\
9CJ0015+2718 & D &&     00:15:58.0   &     27:18:54 &&   0.0031 &   031124 &      \\
9CJ0016+2804 & D &&     00:16:02.4   &     28:04:28 &&   0.0027 &   031124 &      \\
9CJ0016+3158 &   &&     00:16:07.4   &     31:58:53 &&   0.0092 &   041016 &      \\
9CJ0016+3239 &   &&     00:16:11.0   &     32:39:26 &&   0.0230 &   041228 &      \\
9CJ0016+2945 &   &&     00:16:13.0   &     29:45:11 &&   0.0156 &   010815 &      \\
9CJ0016+3238 &   &&     00:16:13.6   &     32:38:37 &&   0.0264 &   041228 &      \\
9CJ0016+3139 &   &&     00:16:40.6   &     31:39:04 &&   0.0110 &   030203 &    b \\
9CJ0017+3209 &   &&     00:17:02.2   &     32:09:19 &&   0.0345 &   041016 &      \\
9CJ0017+2733 & D &&     00:17:04.6   &     27:33:43 &&   0.0062 &   031124 &      \\
\hline

\end{tabular}
\label{table:ctlg}
\end{table*}

A short section of the catalogue is shown in Table~\ref{table:ctlg}. The whole list is available online at http://www.mrao.cam.ac.uk/surveys together with the relevant tables: i.e. Tables \ref{table:areastotal} and \ref{table:areasdeep} from this paper. We have included all 643~sources, which means that a large number of them are below our completeness limits of $\approx10$~mJy for the total areas and $\approx5.5$~mJy for the deeper areas, the faintest being only 1~mJy. Sources which lie within the deeper areas are marked D in the catalogue. In the comment column `b' denotes a binary source and `e' a source which appears extended relative to the synthesized beam size.

We have not assigned individual error estimates to the source parameters. The position of a source is derived from a raster map, unless the follow-up pointed observation indicates a significantly different position, in which case that is substituted. However, since there are large overlaps between adjacent raster maps and a number of sources appear on more than one map, it has been possible to make a general estimate of the position accuracy. Taking the 0020+2947 field, we find 39 sources with repeat position measurements. Apart from one extended source, in no case do the measurements differ by more than 10~arcsec and the median difference is $\sim4$~arcsec. The flux densities are in the range 4.6~mJy to 68.3~mJy. For the brighter sources we can refer to Paper I, where we matched a sample of 17 sources, all $>60$~mJy at 15.2~GHz, with their counterparts in the Jodrell VLA calibrator survey (Wilkinson et al. 1998, and references therein), finding an accuracy for these sources of better than 3~arcsec.

Our flux densities are derived from the pointed observations and the uncertainty in these is dominated by the random uncertainty in the flux calibration, which we estimate to be $\sim5$~per~cent. In the case of an extended source, an estimate is made of the integrated flux density. Since many of the sources are highly variable, the dates of the pointed observations are also included in the catalogue. 

In addition to the catalogues presented in this paper and in Paper I, there are source lists from further areas of the 9C survey yet to be published. We plan to make these available online in due course.

\section{Conclusions}

By developing our techniques of observation and analysis, we have succeeded in extending the 9C survey to a completeness limit of $\approx10$~mJy in an area of 115~deg$^{2}$ and of $\approx5.5$mJy in an area 29~deg$^{2}$. As a blind survey at 15.2~GHz to this depth it provides a unique resource for the study of the extragalactic source population at high radio frequency. It is particularly valuable for the investigation of the foreground point-source contamination in CMB observations, not only for the VSA but for a wide range of other CMB experiments.

\subsection{Source Counts}

In extending our differential source count to 5.5~mJy we detect no evidence of a change from the fitted power law calculated in Paper~I: $n(S) = 51(S/{\rm Jy})^{-2.15}\, {\rm Jy}^{-1}{\rm sr}^{-1}$. We have compared our normalized count $S^{2.5}n(S)$ with that derived from the model of de Zotti et al. (2005) and find that our count appears to lie significantly below their prediction in the lower flux density range.

\subsection{Spectral index distributions}

In discussing the $\alpha_{1.4}^{15.2}$ spectral indices, we distinguish between the properties of a flux-limited sample selected at 15.2~GHz and one selected at 1.4~GHz. The spectral index distributions are necessarily different and the one cannot be inferred from the other in any simple way.

\subsubsection{15.2-GHz samples}

We find that for samples selected at 15.2~GHz in three flux-density bins there is a significant decrease in the median $\alpha_{1.4}^{15.2}$ and increase in the proportion of flat and rising sources with increasing flux density. We cannot identify any observational bias that might produce this effect, but we can compare our results with those of Bolton et al. (2004) who made simultaneous multifrequency observations at 1.4, 4.8, 15.2, 22 and 43~GHz of flux-limited samples of sources selected from 9C at 15.2~GHz. They found that raising the flux limit of the 9C sample produced a significant increase in the fraction of rising-spectrum sources. Their analysis differed from ours in that each of their samples had only a lower flux density limit and no upper limit and also they classified the sources in terms of their 1.4~to~4.8~GHz rather than their 1.4~to~15.2~GHz spectral indices.  However, they also showed that the 1.4~to~4.8~GHz spectral indices from simultaneous measurements were highly correlated with the 1.4~to~15.2~GHz spectral indices obtained by simply matching 9C with NVSS, which is what we have done here. We believe we have observed a significant correlation between spectral index distribution and flux density and that this does appear to be consistent with the work of Bolton et al. (2004).
 
\begin{figure}
        \centerline{\epsfig{file=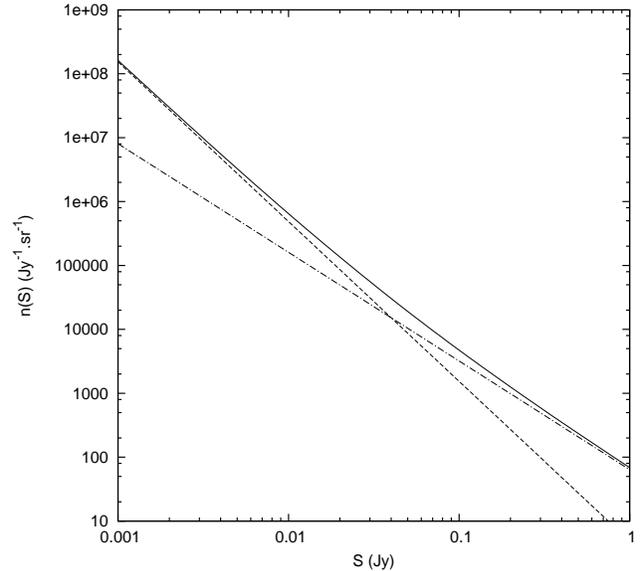,
        angle=270,
        width=11.0cm,clip=}}
\caption{Simulated source counts generated from a model of two populations of radio sources which follow power law source counts. The dot-dashed line corresponds to flat spectrum sources; the dashed line to steep spectrum sources; and the solid line to the total source population for this model.}
\label{fig:sim_counts}
\end{figure}

\begin{figure}
        \centerline{\epsfig{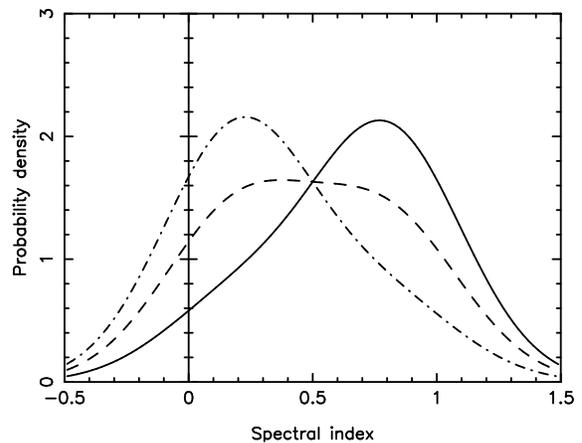}}
\caption{Simulated spectral index distribution generated from a model of two populations of radio sources. The solid line shows the spectral index distribution for sources in the flux density range $5.5 \leq S < 25$~mJy; the dashed line for sources in the flux density range $25\leq S < 100$~mJy; the dot-dashed line for sources $S \ge 100$~mJy. These simulations correspond to the results shown in Figure~\ref{fig:alphas_1}.}
\label{fig:sim_alpha}
\end{figure}

In order to assess whether such a change in spectral index distribution is physically reasonable over such a relatively small range of flux density, we have constructed the following very simplified model. We postulate two different populations of radio sources, one flat spectrum and one steep spectrum. We model these two populations as having differential source counts $\frac{\rm{d}N}{\rm{d}S} \propto S^{\beta}$ with $\beta_{flat} = -1.7$ and $\beta_{steep} = -2.5$. The number counts are normalised such that we have equal numbers of sources at 40 mJy. We generate source counts between 5.5 and 500 mJy as shown in Figure~\ref{fig:sim_counts}; the source count from the combined distribution is consistent with that found from our data.  We then assume that the the flat spectrum sources have spectral indices drawn from a Gaussian distribution with mean $\alpha_{flat} = 0.2$ and $\sigma(\alpha_{flat}) = 0.3$; and that the steep spectrum sources have spectral indices drawn from a Gaussian distribution with mean $\alpha_{steep} = 0.8$ and $\sigma(\alpha_{steep}) = 0.3$.  Using this model we generate the expected spectral index distributions for flux bins $5.5 \leq S < 25$~mJy, $25 \leq S < 100$~mJy and $S \ge 100$~mJy, shown in Figure~\ref{fig:sim_alpha}.

We stress that we have not attempted to optimise this highly simplified model in order to fit our data -- a full physical model of the sort proposed by de Zotti et al. (2005) contains more than two populations and uses a more sophisticated parameterisation of the individual source counts than a simple power law. This model does, however, demonstrate that it is possible to produce a significant shift in the spectral index distribution over a range of two orders of magnitude in flux density due to real astronomical causes rather than being generated by either systematic errors or selection effects. We expect that the spectral index distribution of a sample of sources selected at low frequency, e.g. 1.4 GHz, will show a much smaller variation with flux density owing to the smaller proportion of flat spectrum sources in such a sample.  This expectation appears consistent with the spectral index distribution between 1.4 and 30~GHz, for a sample selected at 1.4~GHz, observed by Mason et al. (2009).

\subsubsection{1.4-GHz samples}

As we have seen, it is not possible to calculate a full spectral index distribution for a sample of sources selected from NVSS at 1.4~GHz but it is useful to explore the flat and rising side of the distribution. We find that, for complete samples of sources above 32.95 and 18.12~mJy at 1.4~GHz, there are 9.8 and 9.2~per~cent respectively with $\alpha_{1.4}^{15.2} \leq 0.5$, and for complete samples above 10 and 5.5~mJy at 1.4~GHz, there are 1.1 and 1.4~per~cent respectively with $\alpha_{1.4}^{15.2} \leq 0$. We note that these values are very close to those found by Mason et al. (2009) for their 1.4 to 31~GHz spectral indices. 

We see that the proportions of sources with $\alpha_{1.4}^{15.2}\leq0$ and $\alpha_{1.4}^{15.2}\leq0.5$ in samples selected at 15.2~GHz (Table~\ref{table:alphas_1}) are significantly higher than those in samples selected at 1.4~GHz (Table~\ref{table:flat_rising}). This effect, of selecting higher proportions of flat-and-rising-spectrum sources the higher the frequency, has been recognised before (see, for example, Peacock~\&~Wall, 1981) and is ascribed to a bias against detecting the increasing number of steep-spectrum sources whose spectra have peaked below the selection frequency and which therefore tend to fall below the flux-density cut-off.

\subsection{Sources not in NVSS}

An important result from our matching with NVSS is the number of sources that we have detected at 15.2~GHz which do not appear in the NVSS catalogue, and, in particular, that in the flux density range of 10 to 15~mJy there are as many as 4.3~per~cent not in NVSS.  This is especially relevant to work that relies upon use of the NVSS catalogue to identify sources at high frequency.  It is very likely to be significant both in the estimation of high frequency source counts and in the selection of individual sources for projecting out of high frequency CMB data.  Our work indicates that the proportion of sources that could be missed by reliance upon the NVSS catalogue might be as high as 4.3~per~cent at low flux densities.

\section{Future work}

We see that all our results are relevant to the estimation of extragalactic source populations at high radio frequencies and of the consequent effects of foreground sources on CMB observations.  In particular, there is the problem of the so-called `CBI excess' (Mason et al. 2003), which is the apparent existence of an excess of power over intrinsic CMB anisotropy on small angular scales ($l > 2000$).  The question of whether this can be ascribed to contamination from the point-source foreground is discussed fully in Mason et al. (2009) and Sievers et al. (2009).

We intend to collaborate with Mason et al. to explore further the comparison of the results in their paper (2009) with ours here and, if possible, to follow up our deeper source list with simultaneous observations at 15 and 30~GHz. Since 9C is so close in frequency this should provide more useful information on the deep source counts at 30~GHz which are critical for interpreting the CBI excess mentioned above.

\section*{Acknowledgments}

We are indebted to the staff of the Mullard Radio Astronomy Observatory for the operation of the Ryle telescope, which was funded by PPARC.  We are grateful to Dave Green for his input to this work and to Sally Hales for managing the online catalogues.  MLD acknowledges an STFC studentship.

\label{lastpage}

\end{document}